\documentclass{jfm}
\usepackage{graphicx}
\usepackage{ulem}
\usepackage{epstopdf, epsfig}
\usepackage{upgreek}
\usepackage{mathtools}
\usepackage{amsmath}

\usepackage{color}

\newcommand{\st}{\sout}

 \newcommand{\be}{\begin{equation}}
 \newcommand{\ee}{\end{equation}}
 \newcommand{\bea}{\begin{eqnarray}}
 \newcommand{\eea}{\end{eqnarray}}

\shorttitle{Droplet dynamics under saturated electrowetting}
\shortauthor{Q. Vo and T. Tran}

\title{Dynamics of droplets under electrowetting effect with voltages exceeding the contact angle saturation threshold}

\author{Quoc Vo\aff{1}
  \corresp{\email{xqvo@ntu.edu.sg}},
 \and Tuan Tran\aff{1}
  \corresp{\email{ttran@ntu.edu.sg}}}

\affiliation{\aff{1}School of Mechanical \& Aerospace Engineering, 
	Nanyang Technological University, 
	50 Nanyang Avenue, 639798, Singapore
}

\begin{document}

\maketitle

\begin{abstract}
Electrowetting-on-dielectric (EWOD) 
is a powerful tool 
in many droplet-manipulation applications 
with a notorious weakness caused by
contact-angle saturation (CAS), 
a phenomenon limiting the
equilibrium contact angle 
of an EWOD-actuated droplet 
at high applied voltage.
In this paper, 
we study the spreading 
behaviours of droplets 
on EWOD substrates 
with the range of applied voltage
exceeding the saturation limit. 
We experimentally find that at the initial stage of spreading,
the driving force
at the contact line 
still follows the Young-Lippmann law
even if the applied voltage is higher than the CAS voltage.
We then theoretically establish
the relation between 
the initial contact-line velocity 
and the applied voltage 
using the force balance at the contact line. 
We also find that 
the amplitude of capillary waves 
on the droplet surface
generated by the contact line's initial motion 
increases with 
the applied voltage.
We provide a working framework 
utilising EWOD with voltages beyond CAS 
by characterising the
capillary waves formed on 
the droplet surface 
and their 
self-similar behaviours.
We finally propose a theoretical model
of the wave profiles
taking into account 
the viscous effects
and verify this model experimentally.
Our results 
provide avenues to
utilise the EWOD effect 
with voltages beyond CAS threshold
and have strong bearing on
emerging applications 
such as digital microfluidic 
and ink-jet printing.
\end{abstract}

\begin{keywords}
Authors should not enter keywords on the manuscript, as these must be chosen by the author during the online submission process and will then be added during the typesetting process (see http://journals.cambridge.org/data/\linebreak[3]relatedlink/jfm-\linebreak[3]keywords.pdf for the full list)
\end{keywords}

\section{Introduction}
\begin{figure}
\centerline{\includegraphics[width=0.8\textwidth]{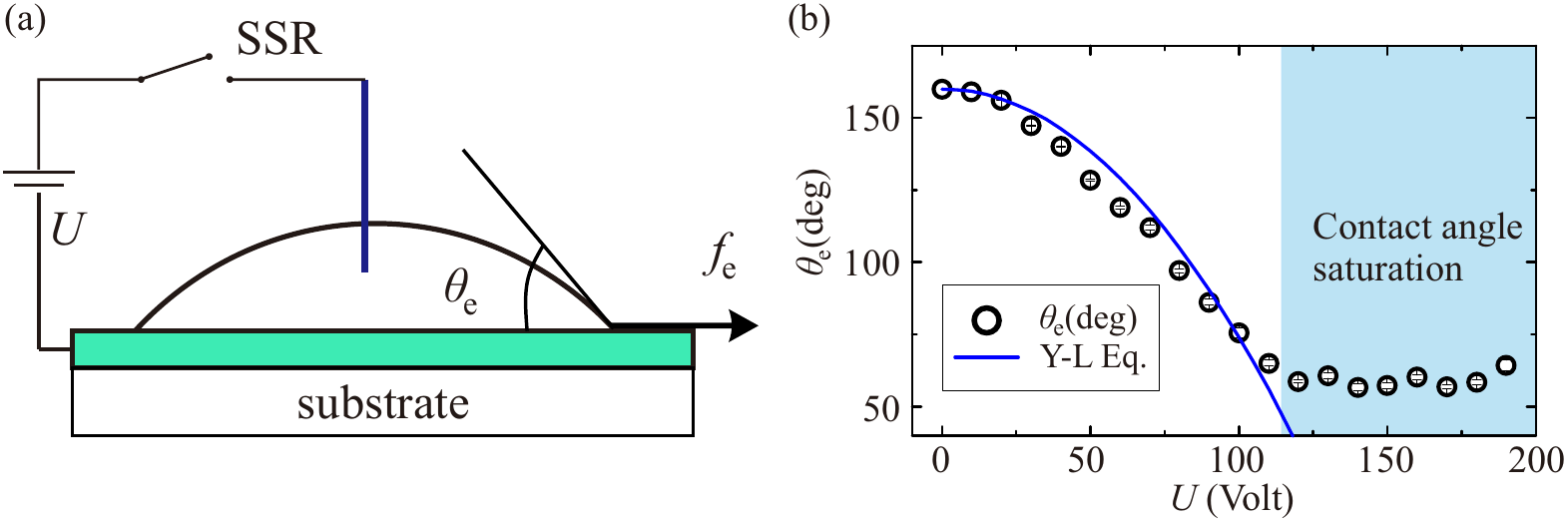}}
\caption{
(a) Schematic showing the
EWOD setup.
(b) Plot showing the dependence 
of the equilibrium contact angle 
$\theta_{\rm e}$ 
on the applied voltage $U$.
The solid curve represents 
Eq.~\ref{eq:YL}
without any fitting parameter. 
Contact angle saturation is observed 
for $U \ge U_{\rm s} \approx 110\,$V.}
\label{fig:principle}
\end{figure}

Contact angle of 
an electrically conductive droplet 
on 
an electrode 
covered by a dielectric layer
reduces when a voltage $U$ 
is applied between the 
droplet and the electrode 
(see Fig.~\ref{fig:principle}a).
This phenomenon, known as electrowetting-on-dielectric (EWOD),
is one of the most effective techniques
for droplet manipulation in
three dimensional digital microfluidics 
\citep{JunLee2012, Hong2015b},
anti-icing
 \citep{Mishchenko2010}, 
 self-cleaning 
 \citep{Blossey2003}
and control of droplet deposition \citep{Baret2006}.
At low applied voltages, 
the equilibrium 
contact angle $\theta_{\rm e}$
after $U$ is applied
directly relates to 
the voltage $U$ 
and droplet's initial 
contact angle $\theta_{\rm 0}$
via the so-called 
Young-Lippmann equation \citep{Mugele2005}:
\begin{equation}
\label{eq:YL}
\cos \theta_{\rm e} - \cos \theta_{\rm 0} = \frac{\epsilon \epsilon_0 U^2}{2 \sigma d},
\end{equation}
where $\epsilon_0$ is 
the permittivity of free space;
$\epsilon$ and $d$ are, respectively, 
the dielectric constant
and thickness  
of the dielectric coating, 
$\sigma$ is surface tension of the droplet.
When the applied voltage is 
sufficiently high, 
$\theta_{\rm e}$
fails to follow Eq.~\ref{eq:YL}
and is limited by
a phenomenon 
known as contact angle saturation (CAS) 
(see Fig.~\ref{fig:principle}b).

While transient behaviours of droplets 
under electrowetting actuation
within the saturation limit
are well-established 
\citep{Mugele2005,Vo2018a,Vo2018},
the 
dynamics of droplets with
applied voltage exceeding
the CAS threshold
remains elusive.
Consequently,
most applications
utilising electrowetting 
uses applied voltage within the saturation value, $U_{\rm s}$,
a requirement significantly 
limiting EWOD's capabilities, 
especially in manipulating 
small and viscous droplets \citep{Vo2019,Fair2007}.
Therefore, it is of practical importance to explore 
how droplet spreading dynamics
is affected when the applied voltage is higher than the 
CAS threshold.

In this paper, 
we investigate
the transient behaviours of 
droplets when the
applied voltage $U$
is higher than the CAS voltage $U_{\rm s}$. 
We focus on 
small droplets 
with radius below capillary length
and study their  
spreading behaviours in silicone oil medium
under electrowetting actuation. 
We reveal a direct influence of increasing
the applied voltage on
enhancing the initial spreading 
velocity of droplets
even when $U \ge U_{\rm s}$.
We also observe strong capillary waves 
on the droplet-oil interface 
and analyse their occurrence 
with the 
enhancement of the initial spreading velocity.
We then propose and verify 
a theoretical model
to explain the relation between 
the capillary waves 
under electrowetting actuation
and the applied voltage. 
 
\section{Experimental setup, materials and method}
In our experiment,
we prepare test substrates 
using indium-tin-oxide (ITO) 
glass slides, 
each covered by a
layer of  
fluoropolymer (Teflon, Dupont) 
having thickness $d = 2.5 {\rm \upmu m}$.
We dip 
an $18\,\mu$m-wire electrode made of tungsten
into the droplet 
and connect it to the positive 
terminal of a DC power supply via
a solid state relay (SSR) 
(see Fig.~\ref{fig:principle}a),
while the negative terminal to the ITO layer.
The 
voltage 
applied to the two electrodes
takes the 
form of 
a step function having amplitude $U$
and 
duration $T_{\rm p}$;
the duration $T_{\rm p}$ is set sufficiently long
to ensure that 
droplets under 
the electrowetting effect
reach new equilibrium
in every actuation.
The amplitude 
$U$ is varied 
between $0\,{\rm V}$ to $190\,{\rm V}$.
The contact angle saturation 
voltage (CAS)  
is experimentally determined at $U_{\rm s} = 110\,$V
by observing the saturation behaviour
of the contact angle 
when $U$ is gradually increased 
(Fig.~\ref{fig:principle}b).
We note that
the observed change of 
the equilibrium contact angle $\theta_{\rm e}$ 
follows the 
Young-Lippmann equation (Eq.~\ref{eq:YL})
when $U < U_{\rm s}$ (Fig.~\ref{fig:principle}b, solid curve).

We use a
0.125\,M sodium chloride
aqueous solution
to generate droplets.
For each experiment, 
we immerse a droplet 
and the substrate in silicone oil 
having viscosity
$\mu_{\rm o} = 1.8\,{\rm mPa s}$.
The interfacial tension  
of the solution in 
the silicone oil 
is measured experimentally at $\sigma = 37.2 \pm 0.5\,{\rm mPa\,s}$.
The temperature of the oil pool 
is kept at $20 \pm 0.5\,^{\circ}{\rm C}$ 
to maintain consistent experimental conditions.
The radius of the droplet
is
$R = 0.5\,{\rm mm}$,
well below
the capillary length 
$\l_{\rm c} = [\sigma/(\rho - \rho_{\rm o}) g]^{1/2} = 5.5\,$mm. 
Here, $g = 9.781\,{\rm m\,s^{-1}}$ 
is the gravitational acceleration,
$\rho = 1000\,{\rm kg\, m^{-3}}$ and
$\rho_{\rm o} = 873\,{\rm kg\, m^{-3}}$ 
the density of the working liquid and the oil, respectively.
We capture the behaviours of droplets
using a high speed camera (Photron, SAX2)
typically running at 5000 frame-per-second.

\section{Results and Discussions}
 \begin{figure}
\centerline{\includegraphics[width=1\textwidth]{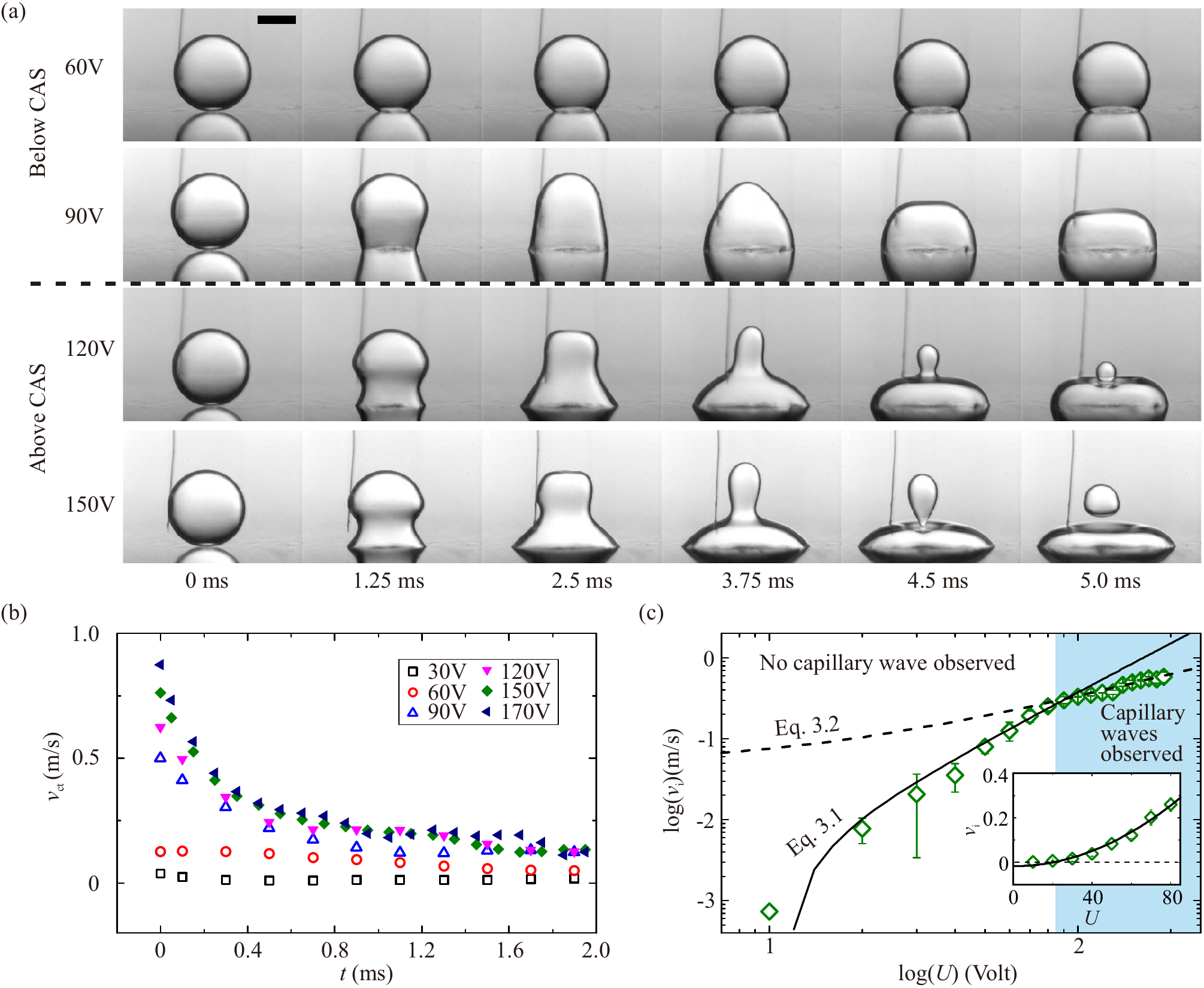}}
\caption{
(a) Snapshots showing the contact-line behaviours of droplets 
under different applied voltages $U$ 
varying
from $60\,{\rm V}$ to $150\,$V. 
The scale bar represents 0.5\,mm.
Capillary waves on droplet's surface 
are observed when 
$U \ge 90\,$V.
(b) Plot showing the
contact-line velocity 
$v_{\rm ct}$ vs. time $t$: 
open markers for those of voltage below the saturation value, i.e., $U < 110\,$V; 
solid markers for those of voltage above the saturation value, i.e., $U > 110\,$V. 
(c) Log-log plot showing 
the initial contact-line velocity 
 $v_{\rm i}$ 
vs. the applied voltage $U$.
Inset: Linear plot of $v_{\rm i}$ vs. $U$ for $U < 90\,$V, the voltage's range in which no capillary wave is observed on the droplet's surface.}
\label{fig:oversaturationDynamics}
\end{figure}

\subsection{Spreading dynamics by electrowetting beyond contact angle saturation}
In Fig.~\ref{fig:oversaturationDynamics}a,
we show several series of snapshots of 
droplets 
shortly after a voltage $U$ is applied.
Generally, we observe that 
higher applied voltage 
causes more violent spreading behaviours
due to
larger contact-line tension imbalance \citep{Mugele2005}.
For instance, at $U = 60\,$V (Fig.~\ref{fig:oversaturationDynamics}a, first panel),
the droplet gently spreads 
and maintains its spherical shape
at all time. 
At $U = 90\,$V (Fig.~\ref{fig:oversaturationDynamics}a, 
second panel),
faster contact-line motion generates
capillary waves on the 
liquid-oil interface and consequently
causes substantial deformation to the droplet's shape.
When the voltage
increases above $120\,$V, 
the capillary waves become 
so strong that 
the deformation at the liquid-oil interface causes
small satellite droplets to
eject from the 
primary droplets 
(Fig.~\ref{fig:oversaturationDynamics}a, two last panels).
In such droplet-ejection instances, 
the size of the satellite droplets 
increases when $U$ varies
from $120\,$V
to $150\,$V.

Surprisingly, 
we also observe from Fig.~\ref{fig:oversaturationDynamics}a that
droplet deformation
increases with the applied voltage
even at $U \ge U_{\rm s} = 110\,$V. 
As the deformation 
is caused by
the capillary waves 
originated from the contact-line
initial motion, 
we hypothesise that
CAS causes little effect to the early dynamics of the contact line.
In Fig.~\ref{fig:oversaturationDynamics}b,
we show 
how the  contact-line velocity $v_{\rm ct}$ 
depends on time
for several values of $U$
from $30$\,V to $170$\,V.
We observe that 
when $U < U_{\rm s}$,
increasing $U$ 
shifts the velocity curves upwards, 
indicating that both early-time 
and late-time spreading
dynamics are affected 
by the applied voltage
(Fig.~\ref{fig:oversaturationDynamics}b, open markers).
However, 
when $U > U_{\rm s}$,
escalating $U$ only increases
the initial values of $v_{\rm ct}$.
Subsequently, $v_{\rm ct}$ 
of all the values of $U > U_{\rm s}$ 
converse 
to the same curve (Fig.~\ref{fig:oversaturationDynamics}b, solid markers).
This observation 
confirms that
the contact-line initial velocity 
and the resulting capillary waves on the droplet's surface are not limited by CAS;
CAS only affects the late-time spreading dynamics of the droplets.

In Fig.~\ref{fig:oversaturationDynamics}c,
we show
the initial contact-line velocity $v_{\rm i}$, 
i.e., 
$v_{\rm ct}$ measured at $t = 0$,
vs. the applied voltage $U$.
We observe that
although $v_{\rm i}$ increases 
with $U$ in the whole range of 
the applied voltage,
the increasing rate of $v_{\rm i}$ 
is higher for $U < 90\,$V
compared to that for $U \ge 90\,$V.
Beyond the critical voltage 
$U_{\rm c} = 90\,$V,
determined where the switch 
in the increasing rate of $v_{\rm i}$
occurs, 
it is possible to observe 
capillary waves on 
the droplet-oil interface
(see Fig.~\ref{fig:oversaturationDynamics}a).
The occurrence of 
the capillary waves 
when the applied voltage 
is increased beyond $U_{\rm c}$
suggests that 
the initial driving force 
generated by 
the electrowetting effect
is opposed by 
the resistive force at the contact line for
$U < U_{\rm c}$,
and by 
the droplet's inertia for 
$U \ge U_{\rm c}$.
As a result, the relation 
between $v_{\rm i}$ and $U$
can be obtained by 
balancing the driving force 
$f_{\rm e} = \epsilon \epsilon_0 (2d)^{-1} U^2$, 
resulted from 
strong localisation
of the electrical field 
at the contact line \citep{Mugele2005},
and either the resitive force
$v_{\rm i} \lambda + f$ for $U < U_{\rm c}$,
or inertia $\rho r_0 v_{\rm i}^2$ for $U \ge U_{\rm c}$.
Here, $\lambda$ is the contact-line 
friction coefficient
described by the geometric mean 
of the droplet's viscosity and the oil's viscosity \citep{Vo2018a}, 
and
$f$ is the contact-line elasticity force 
caused by 
microscopic defects on 
the surface \citep{DeGennes1985,Joanny1984}.
If we assume 
that $f_{\rm e} = \epsilon \epsilon_0 (2d)^{-1} U^2$ is independent from CAS 
in the early-time dynamic, 
we obtain 
\begin{align}
\label{eq:Vm_1}
v_{\rm i} &= \epsilon \epsilon_0 (2 d \lambda)^{-1} U^2 - f \lambda^{-1} 
{}&{\rm for}\,\,U \le U_{\rm c} = 90\,{\rm V},\\
\label{eq:Vm_2}
v_{\rm i} &= (\epsilon \epsilon_0)^{1/2} (2 d \rho r_0)^{-1/2}(U - U_{\rm c}) + v_{\rm c} 
{}&{\rm for}\,\,U \ge U_{\rm c} = 90\,{\rm V},
\end{align}
where
$v_{\rm c} = \epsilon \epsilon_0 (2 d \lambda)^{-1} U_{\rm c}^2 - f \lambda^{-1}$
is the critical contact-line velocity  
above which the capillary wave is 
observed on the droplet-oil interface.
In our experiment,  $v_{\rm c} = 0.3\,{\rm ms^{-1}}$.
In Fig.~\ref{fig:oversaturationDynamics}c,
we plot both Eq.~\ref{eq:Vm_1} 
(solid lines in the main figure and the inset) 
and Eq.~\ref{eq:Vm_2} (dashed line).
We observe an excellent 
agreement 
between the experimental data and
both equations in their respective validity ranges.
Best fit of Eq.~\ref{eq:Vm_1} 
to the data in 
Fig.~\ref{fig:oversaturationDynamics}c 
gives $\lambda = 0.075 \pm 0.005\,{\rm Pa\,s}$, 
in good agreement 
with the previously reported values
for contact-line friction coefficient
in 
the same systems
($\lambda \approx 0.1 \pm 0.01\,{\rm Pa\,s}$) 
 \citep{Vo2018,Vo2018a}.
The fitting also  
reveals the contact-line 
elasticity force 
$f = (2.1 \pm 0.2) 
\times 10^{-3}\,{\rm N m^{-1}}$,
which is of the same order of 
the previously reported value, i.e., 
$(8.6 \pm 0.9) \times 10^{-3}\,{\rm N m^{-1}}$
in \cite{Vo2019}.

\subsection{Capillary waves}
Fast contact line motion generated by the 
electrowetting effect when $U \ge U_{\rm c}$ 
causes capillary waves on the droplet's surface. 
A remarkable feature of the
generated capillary waves is 
that their amplitude
increase with 
the applied voltage $U$
even when $U$ is higher 
than the CAS voltage
(see Fig.~\ref{fig:oversaturationDynamics}). 
To exploit 
this feature in practical applications, e.g.,
manipulating small and viscous droplets \citep{Vo2019,Fair2007}, 
inducing droplet jumping by modulating actuation time \citep{Wang2017}, 
and controlling droplet ejection \citep{Vo2021},
we examine the 
induced capillary waves on the water-oil interface 
to reveal the dependence of the wave amplitude on the applied voltage.

\subsubsection{Capillary-wave generation}
\label{sec:capWaveGeneration}
To induce capillary waves 
on the surface of a droplet, 
the contact-line velocity
has to overcome 
the inertial-capillary velocity 
$(\sigma/\rho r_0)^{1/2}$,  
formulated using
the stabilising velocity of
a deformed droplet's surface  
to minimise its curvature \citep{Taylor1959}.
In 
normal wetting phenomena, 
the maximum velocity
of the contact line
that can be theoretically achieved
is $\sigma/\lambda$.
As a result, 
the condition enabling 
capillary waves becomes 
$\sigma/\lambda \ge (\sigma/\rho r_0)^{1/2}$,
or $\xi = \lambda/(\sigma \rho r_0)^{1/2} \le 1$.
Here, the so-called damping 
coefficient $\xi$
measures the viscous effects in both the droplet and the outer oil 
against inertia for capillary flows. 
We note that $\xi$ is defined 
similarly to the Ohnesorge number 
${\rm Oh} = \mu/(\sigma \rho r_0)^{1/2}$, 
where $\mu$ is the liquid viscosity. 
In wetting phenomena driven by the electrowetting effect, 
$v_{\rm i}$ is determined by Eq.~\ref{eq:Vm_1}
and Eq.~\ref{eq:Vm_2}, 
resulting in
another condition for 
capillary waves to occur: 
$v_{\rm i} \ge (\sigma/\rho r_0)^{1/2}$. 
Using the experimental parameters in our experiment, 
the condition for the contact-line velocity
to generate capillary waves is 
$v_{\rm i} \ge (\sigma/\rho r_0)^{1/2} = 0.27\,{\rm ms^{-1}}$, 
consistent with our experimental results shown in 
Fig.~\ref{fig:oversaturationDynamics}c, 
i.e., 
capillary waves are observable when 
$v_{\rm i} \ge 0.3\,{\rm ms^{-1}}$.
We note that if we define
a Weber-like number
${\rm We} = v_{\rm i}^2 \rho r_0/\sigma$,
the condition for 
capillary waves to occur then
takes the form
${\rm We} \ge 1$.

\begin{figure}
\centerline{\includegraphics[width=0.65\textwidth]{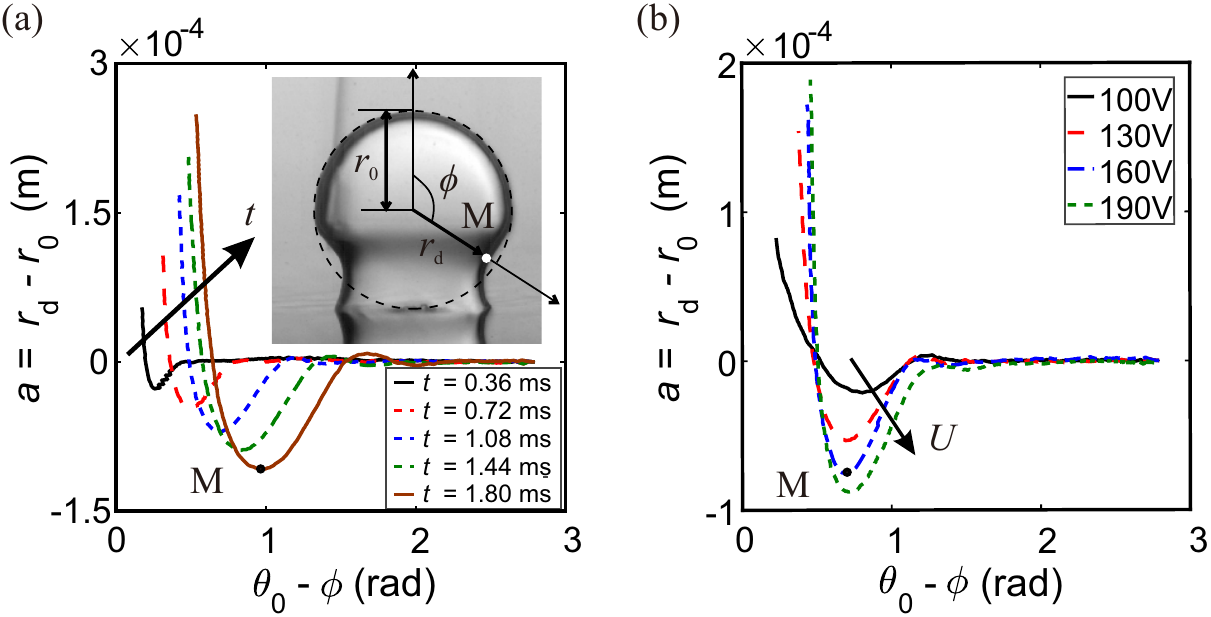}}
\caption{
(a) Plot showing 
the displacement
$a$ vs. 
the wave's position $\theta_0 - \phi$
of the capillary waves 
propagating 
along the droplet-oil interface 
at different time $t$ 
when $U = 150\,$V 
is applied. 
(b) Plot showing $a$ 
vs. $\theta_0 - \phi$
of the capillary waves at  
$t/\tau = 1$ 
and at different 
applied voltages varying from 
$100$\,V to $190$\,V. 
Here, $\tau = (\rho r_0^3/\sigma)^{1/2}$ 
is the inertia-capillary time.
The maximum displacement
(at point M)
increases with 
the applied voltage regardless of
CAS.}
\label{fig:3_capWave}
\end{figure}

\subsubsection{Capillary-wave characteristics}
\label{sec:capWaveCharacteristics}
To quantify the amplitude and phase of 
the capillary waves 
on a droplet having spherical-cap radius
$r_{\rm s}$ and contact angle $\theta_0$,
we use a
polar coordinate system $(r, \phi)$ 
with origin $(r = 0, \phi = 0)$ 
at a distance $r_s$ below the 
droplet's apex
and the 
polar axis vertically upward 
(Fig.~\ref{fig:3_capWave}a, inset).
For droplets in our experiment, 
$\theta_0 \approx \pi$ and 
$r_{\rm s} \approx r_0$. 
In this coordinate system, 
the droplet-oil profile,
which is also 
the capillary waveform, 
is $(r_{\rm d}, \theta_0 - \phi)$,
where $\theta_0$ is 
the wave's position of the contact line 
at $t = 0$.
The displacement of
the capillary wave is therefore
$a = r_{\rm d} - r_s$,
measured by the deviation of 
the droplet-oil interface 
from the droplet's
initial 
shape ($r_d = r_s$).
In Fig.~\ref{fig:3_capWave}a, 
we show 
$a$ vs. $\theta_0 - \phi$
at several instances of time $t$
when $U = 150\,$V is applied.
When $t$ increases, 
the waveform moves from 
left to right
indicating that the 
wave propagates further 
from the contact line and
towards the apex of the droplet 
($\phi = 0$).
Furthermore, we observe that 
while the wave's propagating velocity 
reduces with time,
its amplitude  
increases 
suggesting that the wave's
kinetic energy,
originated from the fast initial 
motion of the contact line, 
is progressively converted
to surface energy
during its propagation.

\begin{figure}
\centerline{\includegraphics[width=1\textwidth]{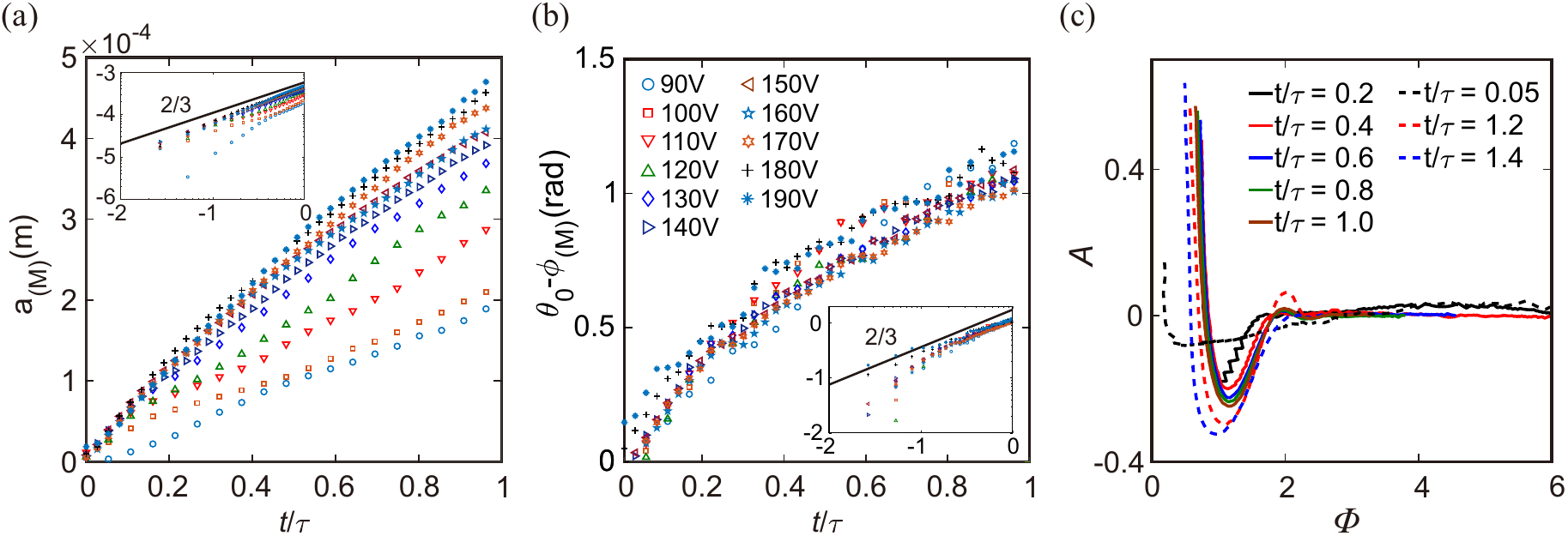}}
\caption{(a) Plot showing the 
displacement
of point M,
$a_{\rm (M)}$, 
vs. $t/\tau$
for $U$ varying from $90$\,V to $190$\,V.
Inset: Log-log plot of the data in 
the main plot showing the 
power law dependence
$a_{\rm (M)}\sim (t/\tau)^{2/3}$.
The legends are given in (b).
(b) Plot showing 
the wave's position of point M,
 $\theta_0 - \phi_{\rm (M)}$, 
vs. $t/\tau$
for $U$ varying from $90$\,V to $190$\,V.
Inset: Log-log plot of 
the data in 
the main plot revealing the 
power law dependence
$\phi_0 - \phi_{\rm (M)} \sim (t/\tau)^{2/3}$.
(c) Plot showing the dimensionless
displacement $A$ 
vs. the dimensionless wave's 
position $\Phi$ 
at different values 
of $t/\tau$ 
varying from $0.05$ to $1.4$. 
We observe self-similar behaviours of 
the capillary waves for 
$0.2 \le t/\tau \le 1$ (solid curves).
Whereas the capillary waves are not self-similar for $t/\tau < 0.2$ and $t/\tau > 1$ (dashed curves).}
\label{fig:4_capWave}
\end{figure}

In Fig.~\ref{fig:3_capWave}b,
we show a plot of
$a$ vs. 
$\theta_0 - \phi$
at $t = \tau$ for 
different values of
$U$ varying from $100$\,V to $190$\,V.
Here, $\tau = (\rho r_0^3/\sigma)^{1/2}$ 
is the inertial-capillary time. 
We observe that 
the displacement
at point M 
significantly 
increases with $U$,
while 
the wave's position of the point M
does not depend on $U$.
This observation is 
examined more quantitatively
in Fig.~\ref{fig:4_capWave}a and b 
where we respectively show
$a_{\rm (M)}$ 
and $\theta_0 - \phi_{\rm (M)}$
vs. $t/\tau$ for 
all the values of 
$U$ ranging from 90\,V to 190\,V.
Here, $a_{\rm (M)}$ and $\theta_0 - \phi_{\rm (M)}$ denote 
the displacement
and the position
of the waves at point M, respectively.
On the one hand, 
we observe that the slopes of 
$a_{\rm (M)}$ vs. $t/\tau$ 
increase with $U$
showing that
the dependence of $a_{\rm (M)}$
on $t/\tau$ is more strongly 
at higher applied voltages.
On the other hand,
$\theta_0 - \phi_{\rm (M)}$ 
is independent from 
$U$ at any values of $t/\tau$
indicating that
the propagating velocity
only depends on 
the hydrodynamical properties 
of the droplets, 
not the contact-line velocity.

The scaling laws 
for 
the displacement
and the position
of capillary waves on 
the surface
of a spreading droplet
were proposed by \cite{Ding2012} by
adapting \cite{Keller1983a}'s self-similarity theory 
for surface-tension driven flows:
\begin{equation}
\label{eq:APhi}
a = A r_0 \left( \frac{t}{\tau} \right)^{2/3}  \,\,{\rm and \,\,}
\theta_0 - \phi = \Phi \left( \frac{t}{\tau} \right)^{2/3},
\end{equation}
where $A$ and $\Phi$ denote the dimensionless 
displacement
and the dimensionless position
of the waves, respectively.
In the insets
of Fig.~\ref{fig:4_capWave}a and b,
we show the log-log plots of $a_{\rm (M)}$ 
vs. $t/\tau$
and $\theta_0 - \phi_{\rm (M)}$ 
vs. $t/\tau$, respectively. 
We observe 
that
$a_{\rm (M)} \sim (t/\tau)^{2/3}$ 
and
$\theta_0 - \phi_{\rm (M)} \sim (t/\tau)^{2/3}$,
implying that 
Eq.~\ref{eq:APhi} 
is also applicable in our system 
in which the capillary waves are generated by
electrowetting actuation.
To further test Eq.~\ref{eq:APhi}, 
in Fig.~\ref{fig:4_capWave}c, 
we show the dimensionless waveforms, i.e.,
$A = a r_0^{-1}(t/\tau)^{-2/3}$
vs. $\Phi = (\theta_0 - \phi)(t/\tau)^{-2/3}$
when $t/\tau$ 
varies from $0.05$ to $1.4$.
We observe that 
all the waveforms 
for $0.2 \le t/\tau \le 1$
collapse 
onto a single one
confirming the self-similar behaviour
of the capillary waves 
in this time range.
However, we notice
that capillary waves 
are not self-similar 
for $t/\tau < 0.2$ and $t/\tau > 1$ (see the dashed curves in Fig.~\ref{fig:4_capWave}c).
At $t/\tau < 0.2$, i.e., 
a short time after wetting
is initiated, 
the waves 
are local at the contact line region \citep{Cox1986a}, 
whereas, at $t/\tau > 1$
the capillary waves reach the apex 
of the droplet
and start interacting with the waves 
coming from the other side of the droplet.
As a result, at both extremes
the capillary waves on the droplet's surface
are more complicated 
and cannot be 
described by a single
self-similar scaling law. 

\subsubsection{Profiles of capillary waves}
With the self-similar behaviour 
of the capillary waves,
it is possible to reduce
the wave's profile $a(\phi,t)$
at any time $t$ 
by a time-invariant dimensionless profile $A(\Phi)$.
The first solution for $A(\Phi)$
was proposed by \cite{Keller1983a} 
who modified the calculation 
of \cite{Jeffreys1999} 
for inviscid, self-similar capillary-driven flows in a fluid wedge.
Later, \cite{Billingham1999} theoretically incorporated 
viscosity effects
using the capillary-viscosity velocity. 
However, both \cite{Keller1983a}'s and 
\cite{Billingham1999}'s solutions 
fail to fit our experimental data. 
While \cite{Keller1983a}'s lacks of 
the viscous effects, 
the model of 
\cite{Billingham1999}
is only applicable 
at a very early time after
the wave is generated,
and in a very small region 
close to the wave's source,
e.g., the time and length scales respectively 
are $\approx 0.2\,$ns 
and $\approx 10\,$nm for water.

\begin{figure}
\centerline{\includegraphics[width=1\textwidth]{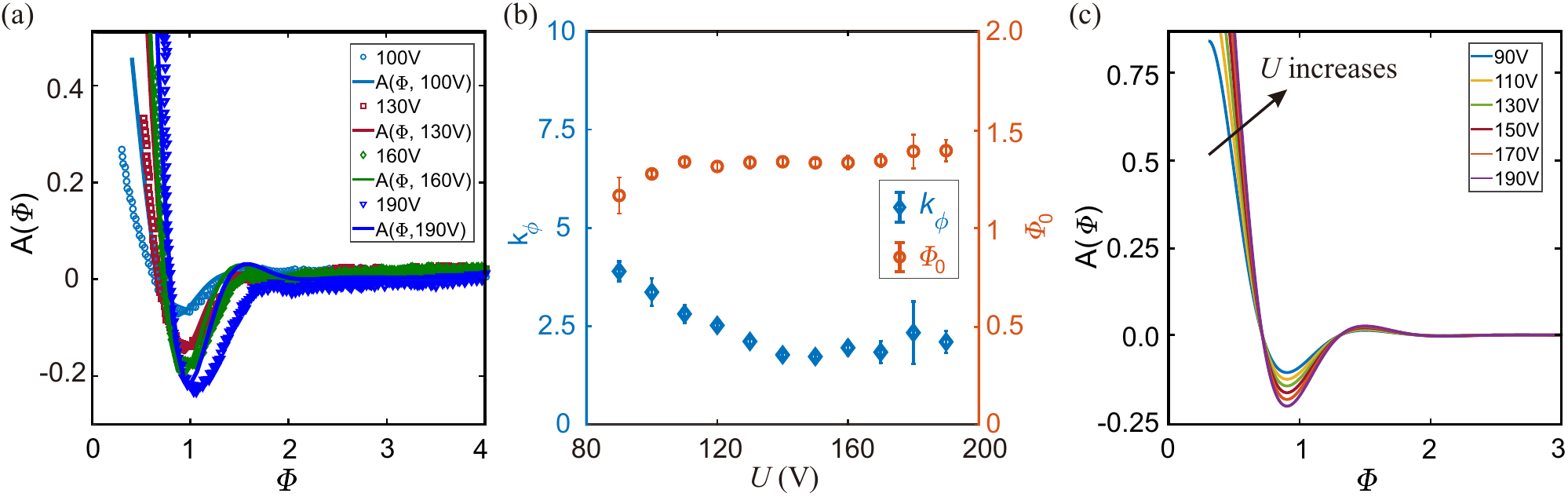}}
\caption{(a) Examples of fitting the experimental data of $A$ vs. $\Phi$
at $t/\tau = 0.25$ 
to Eq.~\ref{eq:waveAmpSolution}. 
Similar results are obtained 
for other values of 
$t/\tau$ varying from $0.2$ to $1$.
(b) Plot showing the fitting parameters $k_\phi$ (on the left axis)
and $\Phi_0$ (on the right axis) 
vs. the applied voltage $U$.
Both $k_\phi = 2.4 \pm 0.7$ 
and $\Phi_0 = 1.32 \pm 0.06$ appear to be constants in the voltage's range.
(c) Plot representing Eq.~\ref{eq:waveAmpSolution} 
for different applied 
voltage varying from $90$\,V to $190$\,V, 
and with $k_\phi = 2.4$ 
and $\Phi_0 = 1.32$.}
\label{fig:capWaveFitting}
\end{figure}

We seek an alternative expression of 
$A(\Phi)$ taking into account the viscosity 
effect 
in our system.
The self-similar profiles $A(\Phi)$
(see the profiles of $0.2 \le t/\tau \le 1$ in Fig.~\ref{fig:4_capWave}c) can 
be approximated by
spatially decaying waves
driven by interfacial tension
and resisted by inertia 
and viscous forces.
For such waves,
we expect that 
$A(\Phi)$ is the solution of 
an one-dimensional time-invariant 
wave equation with a
spatial damping coefficient $\xi = \lambda/(\sigma \rho r_0)^{1/2}$:
\begin{equation}
\label{eq:waveEq}
\frac{d^2 A}{d \Phi^2} + 2 k \xi \frac{d A}{d \Phi} + k^2 A = 0,
\end{equation}
where $k = 2 \pi / \lambda_{\rm \Phi} = 2 \pi$ is the wave number, $\lambda_{\rm \Phi} = 1$
is the dimensionless wavelength.
We note that
the damping coefficient $\xi$ 
does not depend on the strength of the driving force at the contact line.
Our discussion is also limited 
to the case 
that $\xi < 1$
for the possibility of capillary wave generation 
as discussed in 
Section.~\ref{sec:capWaveGeneration}.

If we denote $\Phi_0 \ge 0$ the 
dimensionless position 
of the contact line,
the profile $A$ in Eq.~\ref{eq:waveEq}
is required to satisfy the 
following boundary conditions:
\begin{align}
\label{eq:BC1}
A\bigg \rvert_{\Phi = \Phi_0} &= A_{\rm ct} = \frac{a _{\rm ct}}{r_0 (t/\tau)^{2/3}},\\
\label{eq:BC2_1}
\frac{d A}{d \Phi}\bigg \rvert_{\Phi = \Phi_0} 
&= \frac{d A_{\rm ct}}{d \Phi_{\rm ct}}
= \frac{\dot{a}_{\rm ct} r_0^{-1}  - (2/3) a _{\rm ct} r_0^{-1} t^{-1}}{- \dot{\phi}_{\rm ct} - (2/3) (\theta_0 -  \phi_{\rm ct})t^{-1}},
\end{align}
where $a _{\rm ct} = r_{\rm ct} (\sin \phi_{\rm ct})^{-1} - r_{\rm 0}$ 
is the contact line's 
displacement, 
$\phi_{\rm ct}$ its wave's position,
$r_{\rm ct}$ the contact radius, 
$\dot{a}_{\rm ct}$ 
denotes the time derivatives of $a_{\rm ct}$
and $\phi_{\rm ct}$ the time derivatives of $\dot{\phi}_{\rm ct}$.
As we consider the capillary waves 
generated short-time 
after the electrowetting effect 
is activated,
we can approximate 
$a _{\rm ct} \approx 0$,
$\phi_{\rm ct} \approx \theta_0$,
$a _{\rm ct} t^{-1} \approx \dot{a}_{\rm ct} \approx v_{\rm i} \sin \theta_0$
and $- (\theta_0 -  \phi_{\rm ct})t^{-1} \approx \dot{\phi}_{\rm ct} = - k_{\phi} r_0^{-1} (\sigma/\rho r_0)^{1/2}$,
where $k_{\phi}$ 
is a positive constant. 
As a result, we obtain
\begin{align}
\label{eq:BC_1}
A\bigg \rvert_{\Phi = \Phi_0} &= 0,\\
\label{eq:BC_2}
\frac{d A}{d \Phi}\bigg \rvert_{\Phi = \Phi_0} 
&= \frac{v_{\rm i} \sin \theta_0}{k_{\phi} (\sigma/\rho r_0)^{1/2}}
= \frac{{\rm We}^{1/2} \sin \theta_0}{k_{\phi}}.
\end{align}

The general solution of Eq.~\ref{eq:waveEq} is 
\begin{equation}
\label{eq:waveAmp}
A = {\rm e}^{-k \xi (\Phi - \Phi_0)} \left[\alpha \cos k_{\rm d} (\Phi - \Phi_0) + \beta \sin k_{\rm d} (\Phi - \Phi_0)\right].
\end{equation}
Here, $k_{\rm d} = 2 \pi (1 - \xi^2)^{1/2}$ 
is the attenuated wave number due to 
the damping coefficient $\xi$.
Applying Eq.~\ref{eq:BC_1} 
and Eq.~\ref{eq:BC_2} 
to Eq.~\ref{eq:waveAmp}, 
we obtain $\alpha = 0$ and
$\beta = {\rm We}^{1/2} \sin \theta_0 (k_{\rm d}  k_{\phi})^{-1}$.
Consequently, we obtain
an explicit expression of $A(\Phi)$:
\begin{equation}
\label{eq:waveAmpSolution}
A = \frac{{\rm We}^{1/2} \sin \theta_0}{k_{\phi} k_{\rm d}}\, {\rm e}^{-k \xi (\Phi - \Phi_0)}\,\sin [k_{\rm d}(\Phi - \Phi_0)].
\end{equation}

To verify Eq.~\ref{eq:waveAmpSolution}, 
we use the least-mean-square method 
to fit the experimental data of
$A$ 
vs. $\Phi$ 
to Eq.~\ref{eq:waveAmpSolution} 
with the
fitting parameters
$\Phi_0$ and $k_\phi$. 
Examples of the fittings 
at $t/\tau = 0.25$ and
for different applied voltages $U$ 
are shown in Fig.~\ref{fig:capWaveFitting}a.
Similar results 
obtained for different values of $t/\tau$ varying from $0.2$ to $1$
indicate a good 
agreement between 
Eq.~\ref{eq:waveAmpSolution} 
and the experimental data. 
In Fig.~\ref{fig:capWaveFitting}b, 
we show 
the values of $k_\phi$ (left axis) 
and $\Phi_0$ (right axis)
that both can be approximated as constants, i.e., 
$k_\phi = 2.4 \pm 0.7$ 
and $\Phi_0 = 1.32 \pm 0.06$, 
for 
$U$ varying from 90\,V to 190\,V.
All solutions for $90 \le U \le 190$\,V
are plotted in 
Fig.~\ref{fig:capWaveFitting}c.

\section{Conclusions}
We have shown that the initial
contact-line velocity 
of a droplet 
spreading on a solid substrate under 
the electrowetting effect
is not affected by the 
contact angle saturation (CAS).
We theoretically derived and experimentally verified
the relation between the initial contact-line velocity
and the applied voltage.
The magnitude of the capillary waves on 
the droplet-oil interface also
depends on the applied voltage
regardless of CAS.
Based on 
theories describing the 
self-similar behaviours 
of interfacial-tension-driven flows, 
we then proposed a 
mathematical expression and provided experimental verification
for the far-field 
capillary-wave profiles incorporating viscous effects. 
The proposed model strengthens 
our current
understanding on 
how the capillary waves generates,
propagates and decays.
The model is also useful  
in understanding other capillary-wave-driven
phenomena such 
as droplet jumping \citep{Vahabi2018}, 
droplet/bubble ejection 
by coalescence \citep{Zhang2015b, Zhang2008,Zhang2009}, 
bubble-busting jet at liquid-air interface \citep{Gordillo2019b},
and bouncing of impacting droplets 
on a solid substrate \citep{Richard2002,Renardy2003}.

\section*{Acknowledgments}
This study is supported by Nanyang Technological
University, the Republic of Singapore's 
Ministry of Education (MOE, Grant No. MOE2018-T2-2-113),
and the Agency for Science, 
Technology and Research (A*STAR, Grant No. 1523700102).

\section*{Declaration of Interests}
The authors report no conflict of interest.

\bibliographystyle{jfm}

\end{document}